# Near-Unity Absorption in Van der Waals Semiconductors for Ultrathin Optoelectronics

Deep Jariwala,[1,2] Artur R. Davoyan,[1,2,3] Giulia Tagliabue,[1,4] Michelle C. Sherrott,[1,2] Joeson Wong[1] and Harry A. Atwater[1,2,3,4]*

[1]Department of Applied Physics and Materials Science, California Institute of Technology, Pasadena, CA-91125, USA

[2]Resnick Sustainability Institute, California Institute of Technology, Pasadena, CA-91125, USA

[3]Kavli Nanoscience Institute, California Institute of Technology, Pasadena, CA-91125, USA

[4]Joint Center for Artificial Photosynthesis, California Institute of Technology, Pasadena, CA-91125, USA

*Corresponding author: haa@caltech.edu

Abstract:

We demonstrate near unity, broadband absorbing optoelectronic devices using sub-15 nm thick transition metal dichalcogenides (TMDCs) of molybdenum and tungsten as van der Waals semiconductor active layers. Specifically, we report that near-unity light absorption is possible in extremely thin (< 15 nm) Van der Waals semiconductor structures by coupling to strongly damped optical modes of semiconductor/metal heterostructures. We further fabricate Schottky junction devices using these highly absorbing heterostructures and characterize their optoelectronic performance. Our work addresses one of the key criteria to enable TMDCs as potential candidates to achieve high optoelectronic efficiency.

TOC:

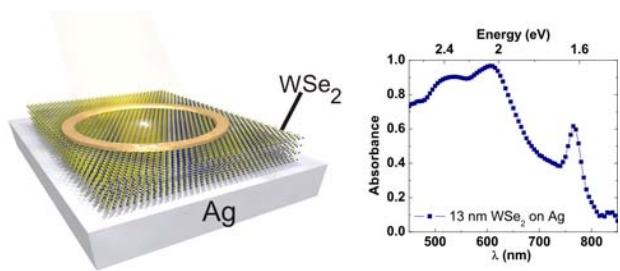

Advances in synthesis, processing and nanofabrication of low-dimensional materials over the last two decades have enabled significant progress towards thin semiconductor layers for high efficiency optoelectronics[1-4] and for solar energy conversion applications.[5-8] For established crystalline inorganic semiconductor absorbers, light management structures such as microwire arrays,[9, 10], Mie resonators[11], photonic crystals[12, 13] and plasmonic metal nanostructures[14, 15] enable enhanced absorption in the active layers, and reduced reflection. In conventional crystalline semiconductors, achieving the necessary surface passivation while incorporating such light management structures is a considerable challenge, since an increasing surface/volume ratio typically results in reduced radiative efficiency. The emergence of two dimensional (2D) semiconducting atomic layers namely TMDCs of molybdenum and tungsten[16] has opened up a new class of high radiative efficiency semiconductors that can be synthesized in ultrathin form. Several reports have demonstrated the use of TMDCs as active layers in optoelectronic and photovoltaic devices. Most reports have utilized TMDCs in a back-gated van der Waals Schottky junction geometry with graphene[17, 18], a van der Waals p-n heterojunction[19, 20] or in an electrostatically split-gated p-n homojunction[19, 21] geometry. In spite of recent theoretical and experimental advances in light trapping in ultrathin 2D layers, [22-26] in most approaches to date, the absorption in active layer is far from optimal and often narrowband or highly sensitive to the angle of incidence.

Metallic rear surfaces are commonly used for enhancing light absorption in optoelectronic devices. For structures whose thickness is greater than wavelength scale, the performance of the metallic rear surface can be interpreted as a simple ray optical specular reflector. However when the semiconductor absorber/reflector heterostructure thickness is at or below the wavelength scale, a different conceptual approach is needed. Prior computational investigations have shown that thin absorber/metal heterostructures result in light absorption enhancement due to an increase in the local density of states (LDOS) near the semiconductor/metal interface.[27, 28] If the heterostructure is thin, then light absorption can be enhanced in a broadband manner, corresponding to enhanced absorption close to the interface, when a thin semiconductor is placed in intimate planar contact on a reflecting metal substrate.[27] This concept was then demonstrated experimentally in ultrathin (< 25 nm) germanium (Ge) on gold (Au) and silver (Ag).[29] However it is difficult to thin down covalently bonded, isotropic 3D semiconductors to below 100 nm thickness without significant degradation of crystalline quality, increasing defect density or influence of surface oxides and states on electronic charge transport. This imposes limitations on the applicability of 3D semiconductors in ultrathin photovoltaic devices. By contrast, TMDCs have self-passivated, dangling bond- and oxide-free surfaces[16, 30] and are thus attractive alternatives for ultrathin absorbers when coupled with reflective metals (Figure 1 a). Here we report near-unity, broadband absorption in ultrathin (12-15 nm) TMDC layers and demonstrate proof-of-concept devices as potential candidates for photovoltaic applications.

An initial look at a micromechanically exfoliated $WSe_2$ structure on a template stripped[31] Ag substrate in broadband white-light illumination shows regions of stark and varying color contrasts from pale red to dark blue (Figure 1b). Observing at higher magnification further reveals crystalline flakes with uniform smooth, straight edges, stepped layers and thickness variations akin to numerous prior observations of exfoliated 2D crystals on $SiO_2$ substrates.[32, 33] Measurement of thickness with atomic force microscopy (AFM) indicates flake thicknesses varying from ~3 nm (pale red) to 13 nm (nearly black) (Figure 1 c-e) suggesting a highly absorbing nature. The step height and surface roughness (root mean square roughness < 1 nm for Ag and < 0.3 nm for WSe$_2$) are also highly uniform as seen in the AFM topography (Figure 1 f).

Absorption spectra for varying thickness WSe$_2$ on Ag (Figure 2 a) back reflector were calculated using available values of refractive index and extinction coefficient from the literature[34] to quantify the above observations. Strongly enhanced absorption was observed with increasing thickness of the WSe$_2$ with near-unity absorption peak occurring between 500-650 nm for varying thickness of the flakes. The peaks in absorption, except for the primary exciton peak at the absorption edge, undergo a red shift with increasing thickness of WSe$_2$ suggesting dependence on the optical path length implying thin film interference effect where the reflected light is strongly attenuated, leading to non-trivial interface phase shifts.[29] Briefly, in the case of a perfect metal/lossless dielectric ($\kappa$=0) with refractive index n the phase shift at the metal dielectric interface is $\pi$ corresponding to perfect reflection. Hence a minimum dielectric film thickness of $\lambda/4n$ on the metal would form an optical cavity with 0 or $2\pi$ phase shift at the dielectric/air interface. If the dielectric is lossy ($\kappa \neq 0$) however, even for thicknesses in the deep subwavelength regime, the total reflection and transmission phase shifts can be approximately 0 or $2\pi$ at the air/dielectric interface giving rise to an absorbance resonance as seen in Figure 1 c-e. Experimentally acquired spectra of WSe$_2$ on template stripped Ag surfaces (Figure 2b) shows remarkably good qualitative and quantitative agreement with the calculations. An interesting observation in the above experiments is that a broadband perfect absorption only occurs for a narrow range of WSe$_2$ thicknesses between 12-15 nm only with an intimate contact with a metal back reflector (See Supporting information S1 for more details). Below or above this thickness, there is increased reflection in the red or blue parts of the spectrum leading to net reduction in integrated absorption. Further, for bulk free standing or glass supported TMDCs, the maximum above-gap absorption is limited to a maximum of ~40%. Due to the large index mismatch, a large fraction (50-60%) of the incident light is reflected back from the surface of bulk crystals[35, 36] (See supporting information figure S1). Likewise, the absorption in few layer-bulk TMDCs on the conventionally used Si/SiO$_2$ substrates is also limited to a maximum between 50-60%[37, 38] The above observations are not unique to WSe$_2$ and can be further generalized to other TMDCs (Figure 2 c-f) as well as Au back reflectors (See Supporting Information S2).

Although the absorption peaks in our structure are dependent on path length, they are highly insensitive to the angle of incidence as a can be seen for the case of 13 nm WSe$_2$ on Ag (Figure 3a). The peak absorption stays over 80% even at a 60° incident angle (Figure 3 b) suggesting relatively low sensitivity to the angle of incident light. This feature of TMDC/Ag heterostructures is highly advantageous for off-normal light collection and may be of a particular interest for photovoltaic applications and solar energy harvesting.[9, 39]

Based on the above discussion, it is evident that the TMDC/metal stack is a suitable ultrathin absorber for a light-harvesting device. To demonstrate this concept, we fabricated a simple device, as shown in Fig. 3a with a metal ring electrode on top using standard photolithography and metal evaporation. The back reflector combined with a patterned metal electrode on top of the flake creates a metal$_1$/TMDC/metal$_2$ sandwich structure (Figure 4 a-b) that can effectively function as a Schottky barrier device if there is sufficient difference between work functions of metal$_1$ and metal$_2$ (Figure 4c). Considering the small size of the top ring electrode and a conductive metallic back substrate, the devices can only be probed accurately while being viewed under a high magnification (50 x), long working distance objective. Upon broadband, white light illumination, (Hg vapor lamp, X-Cite 120 Q) the devices show a pronounced photovoltaic response (Figure 4d). To deduce the collection area and current density, a spatial photocurrent map is acquired using scanning photocurrent microscopy (Figure 4e). The photoexcited carriers diffuse and get collected from approximately 1-3 µm region in the vicinity of the inner and outer metal ring contact boundary (see Supporting information S3). Based on this, we estimate photocurrent density values in Figure 4f. While, the incident light on the device is focused owing to the nature of the measurement and the small size of the device, it is still noteworthy that for ~20 x concentrations (2.1 W/cm$^2$), the short circuit current density ($J_{SC}$) is > 10 mA/cm$^2$. Considering that semiconducting TMDCs are still in the early research phase in terms of material quality and crystal defect control, these photocurrent values are promising in an un-optimized device structure. The van der Waals interlayer bonding in TMDCs induces some level of electron-hole confinement at all thicknesses. Thus, exciton binding energies even in bulk TMDCs are ~70-80 meV.[40] To investigate if the photocurrent is limited by lack of exciton dissociation or free carrier recombination, the exponential dependence of photocurrent on incident light intensity was investigated (Figure 4f, inset). An exponent close to unity points to monomolecular recombination[41] suggesting excitons recombining at neutral impurity or one of free carriers reacting with an oppositely charged impurity.

Finally, we investigate the spectral dependence of photocurrent by illuminating with a laser focused on a fixed spot generating photocurrent in a 12 nm WS$_2$/Ag device (Figure 5 a). For input powers of 1.6 µW at 633 nm corresponding to the primary exciton peak of WS$_2$, we observe pronounced photovoltaic effect with open circuit voltages ($V_{OC}$) approaching 0.2 V and $I_{SC}$ > 100 nA, resulting in a single-wavelength power conversion efficiency ~ 0.5 % (Figure 5b). At this power, the external quantum efficiency (EQE) is ~ 13 % comparable with previously reported values in multilayer devices. At higher input power, the efficiency drops down to below 8% (Figure 5c) suggesting increasing recombination with increasing carrier density, indicating a carrier density dependent recombination mechanism such as Auger recombination. The EQE also remains relatively constant between 8-12 % above the absorption edge as seen in WSe$_2$ on Au (Figure 5d) and its spectrum roughly corresponds to the absorption one. The resulting above-gap IQE is a modest 10% across the absorption spectrum. The lack of high quantum efficiency can be attributed to several factors. Primary among them is the device geometry which prohibits optical excitation of the TMDC directly beneath the top metal electrode which results in in-plane diffusion of carriers for collection. Second, the Schottky junction leads to recombination of all excitons and free electron hole pairs that reach the metal electrode. Finally, the semiconductor quality remains far from optimal as evidenced from the exponent of power dependence of photocurrent suggesting monomolecular recombination. The carrier collection and EQE may be improved by use of transparent top contact such as graphene[17, 18] in addition to a type-II heterojunction between two TMDCs[42] (See Supporting Information S5).

In summary, we have shown an ultrathin, near-unity, broadband semiconducting absorber system using TMDC/metal heterostructure and have applied it in Schottky junction optoelectronic devices. It is also worth noting that most light trapping techniques in thin optoelectronics involve the integration of a patterned nanostructure which could significantly add to the total cost and complexity of the resulting device. In contrast, the above presented results avoid the use of any nanopatterning to enhance light absorption. With further development of the presented structure to introduce a p-n junction and carrier selective contact layers, we expect that it might be possible to engineer $V_{OC} > 1$ V and thus eventually obtain meaningful power conversion efficiencies. The efficient light absorption results reported here, combined with the recent demonstration of near-unity luminescence quantum yield in $MoS_2$,[43] and advances to improve the TMDC material quality[44] hold promise for future high-efficiency, ultrathin optoelectonics and photovoltaics with TMDC active layers.

**METHODS:**

**Sample preparation.** TMDC flakes were deposited on template stripped Au and Ag via mechanical exfoliation of bulk crystals (HQ Graphene). The resulting flakes were identified by optical microscopy and later characterized by AFM to determine the flake thickness. The Au and Ag films were deposited by electron beam and thermal evaporation respectively without any adhesion layers on Si wafers with native oxide only. Standard solvent and plasma cleaning procedures were used for cleaning Si wafers prior to deposition. The substrate was heated to 100 ˚C during thermal evaporation of Ag and the deposition rates were maintained at ~0.1 Å/sec for the first 30 nm in the case of both Au and Ag followed by ramping up to ~1 Å /sec till the final thickness reached 120 nm. The metal films were then template stripped using a thermal epoxy (Epo-Tek 375, Epoxy Technology) using a procedure described in ref.[31]

**Device fabrication, absorbance and photocurrent measurements.** Devices were fabricated using standard photolithography and thermal or e-beam metal evaporation. All absorbance measurements and the EQE spectrum measurements were performed using a home built absorption measurement setup. Tunable, monochromatic light (400-1800 nm) was obtained by coupling a supercontinuum laser (Fianium) to a monochromator. The collimated, monochromatic beam, was then focused on the sample with a long working distance (NA = 0.55), 50x objective and the reflection was measured with a Si detector. The used objective ensures close-to-normal incidence illumination of the device. The reflection spectrum was then normalized to the reflections from a silver mirror (Thorlabs). In the absence of transmission, absorption was obtained as 1-Normalized Reflection (see Supporting information S4). Electrical measurements were performed using Keithley 2400 and 236 source meters and custom LabView programs. The spatially varying photocurrent measurements and global broadband illumination measurements were performed on a scanning confocal microscope (Zeiss, LSM 710) and the incident laser power was measured using power meter (ThorLabs). The devices were probed using piezo controlled microbot manipulators (Imina Technologies) and all measurements were performed under ambient temperature and pressure conditions.

## ASSOCIATED CONTENT:

**Supporting Information**:

Experimental methods, additional experimental data, calculations and analysis accompany this paper. This material is available free of charge via the Internet at http://pubs.acs.org

## AUTHOR INFORMATION

**Corresponding authors:**

*Harry A. Atwater, E-mail: haa@caltech.edu

## NOTES:

**Competing financial interests**: The authors declare no competing financial interests.

## ACKNOWLEDGEMENTS

This work is part of the 'Light-Material Interactions in Energy Conversion' Energy Frontier Research Center funded by the U.S. Department of Energy, Office of Science, Office of Basic Energy Sciences under Award Number DE-SC0001293. D.J., A.R.D and M.C.S. acknowledge additional support from Resnick Sustainability Institute Graduate and Postdoctoral Fellowships. A.R.D also acknowledges support in part from the Kavli Nanoscience Institute Postdoctoral Fellowship. G.T. acknowledges support in part from the Swiss National Science Foundation, Early Postdoc Mobility Fellowship n. P2EZP2_159101. J.W. acknowledges support from the National Science Foundation Graduate Research Fellowship under Grant No. 1144469.

**Author contributions:** D.J. prepared the samples and fabricated the devices. A.R.D. performed all the calculations. D.J., G.T. and J.W. performed the electrical and photocurrent measurements. M.C.S. assisted with sample preparation and fabrication. H.A.A. supervised over all the experiments, calculations and data collection. All authors contributed to data interpretation, presentation and writing of the manuscript.

**FIGURES:**

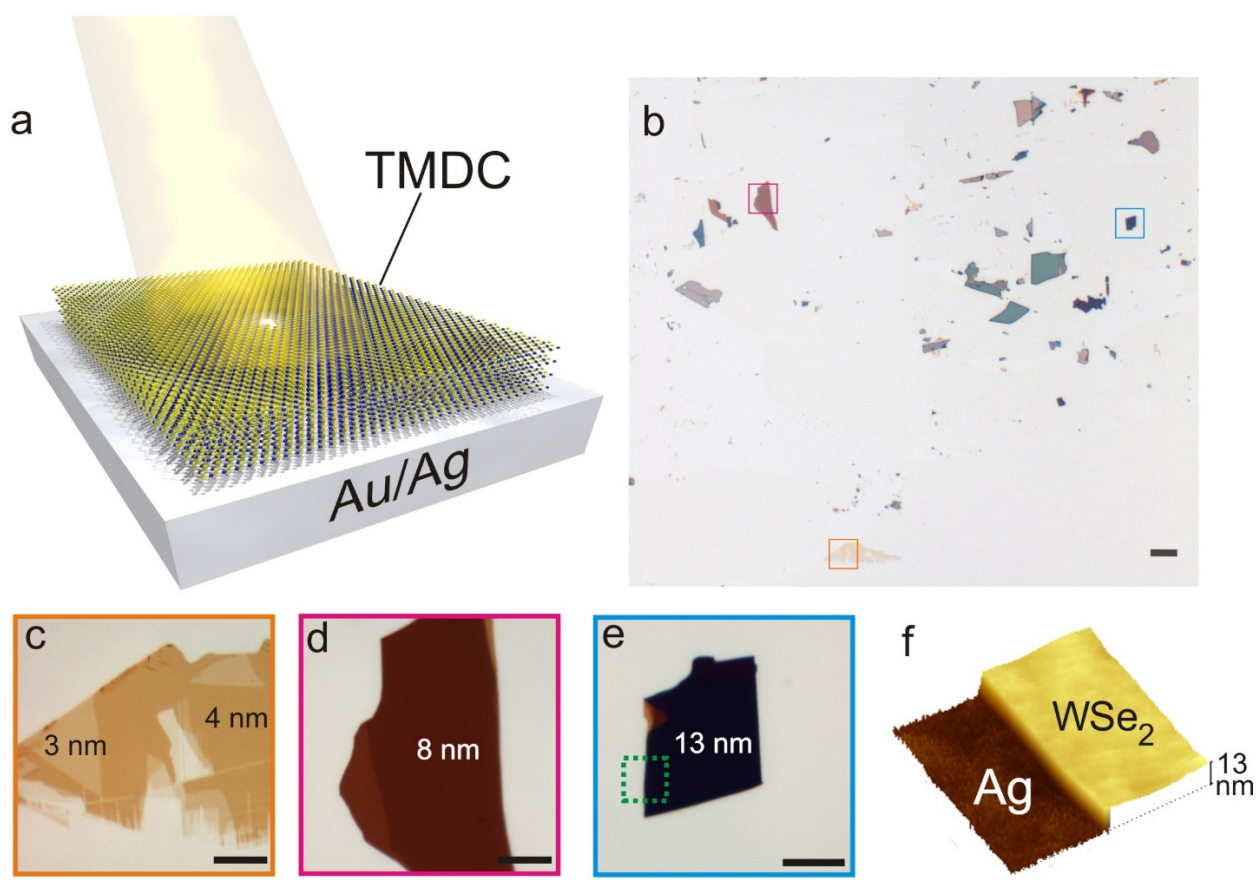

**Figure 1. Absorbing dielectrics on metals: a.** Schematic diagram of a thin, multilayer TMDC film on a Au/Ag back reflecting substrate. **b.** Low magnification optical micrograph of exfoliated WSe$_2$ flakes of on template stripped Ag substrate. (Scale bar = 50 μm) **c-e.** High magnification

micrographs of yellow, red and blue square regions on **(b)** respectively with increasing flake thickness from **(c)** to **(e)**. The sharp blue shift in color and rising contrast with increasing thickness can be seen (Scale bar = 10 μm). **f.** AFM topography of the flake region in (**e**). denoted by the green dashed square.

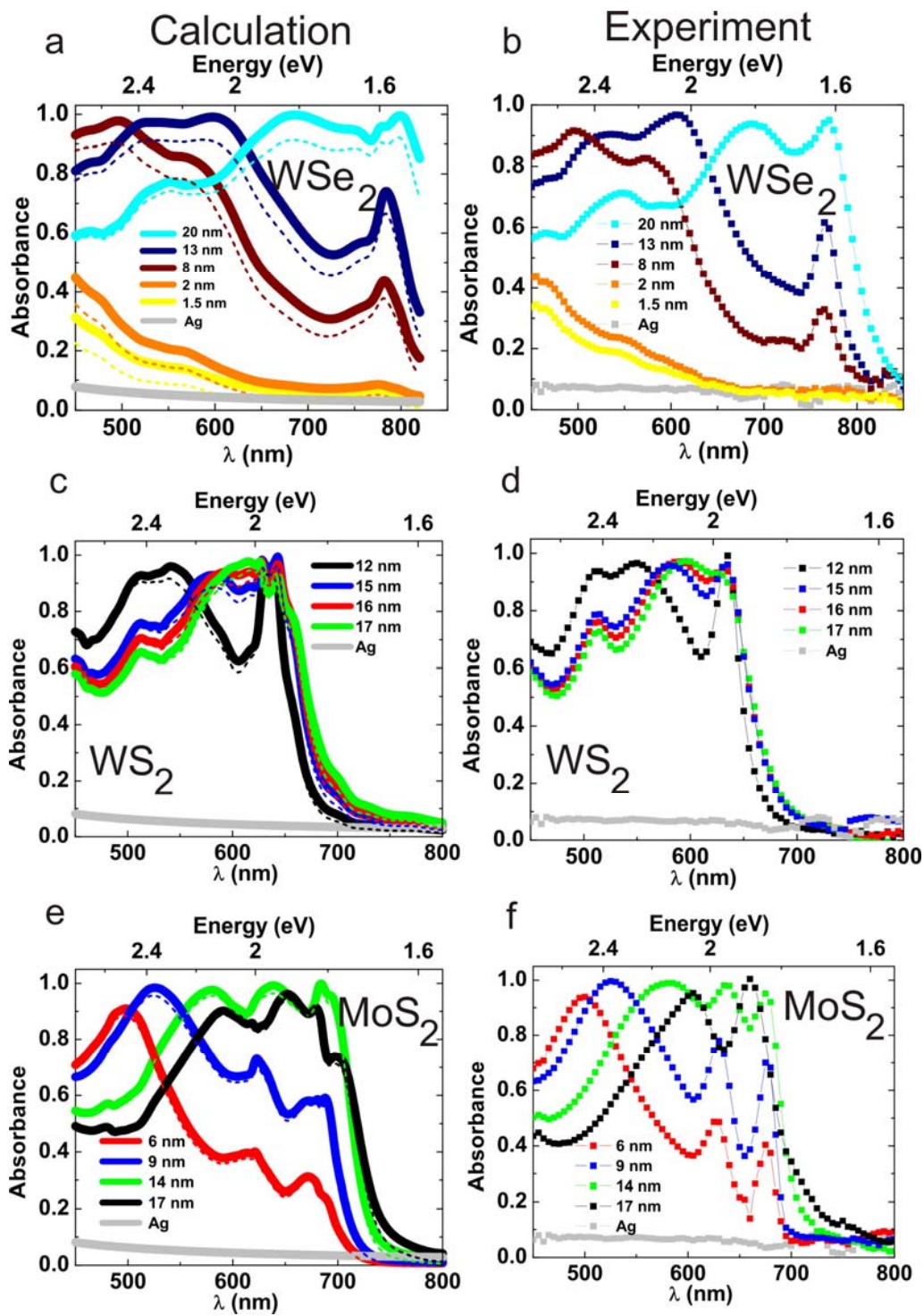

**Figure 2. Absorption spectra of ultrathin TMDCs on Ag back reflector: a.** Calculated absorption spectra of varying thicknesses of WSe₂ on an optically thick Ag film. The solid lines represent total absorption in the WSe₂/Ag stack while the dashed lines represent absorption only in the WSe$_2$. **b.** Experimentally measured absorption spectra of WSe$_2$ flakes exfoliated on template stripped Ag films. **c-d.** Same as a-b except for WS$_2$ on Ag. e-f Calculated (e) and measured (f) absorption spectra for varying thicknesses of MoS$_2$ on Ag.

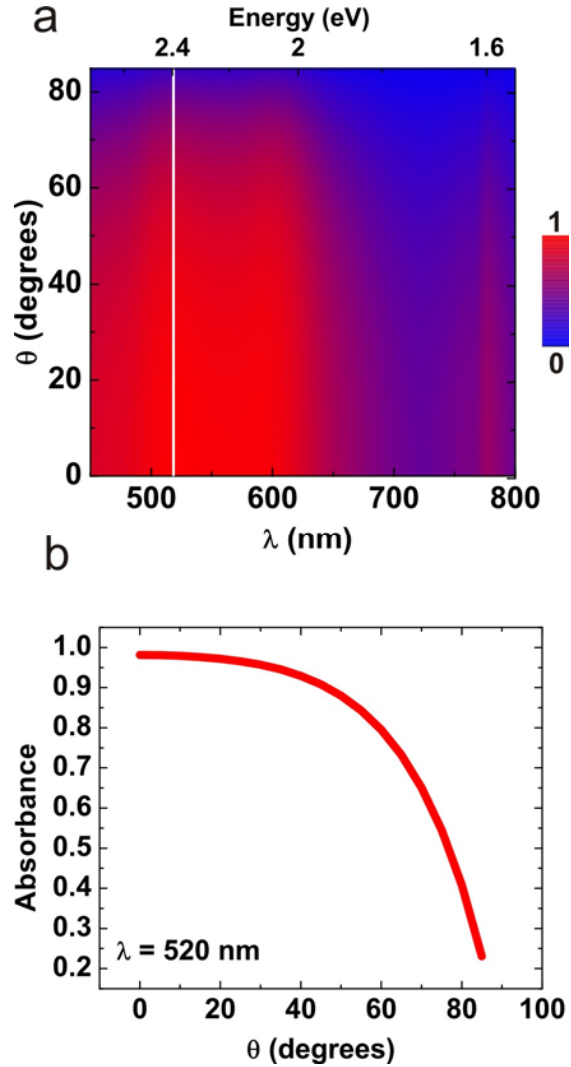

**Figure 3. Angle dependence of absorption in TMDC/Ag heterostructures: a.** Contour plot of calculated absorption spectra at varying angles for 13 nm WSe$_2$ on Ag back reflector. The insensitivity of the absorption as a function of incident angle is apparent. **b.** Line cut from **(a)** at 520 nm showing the angle dependence of peak absorption.

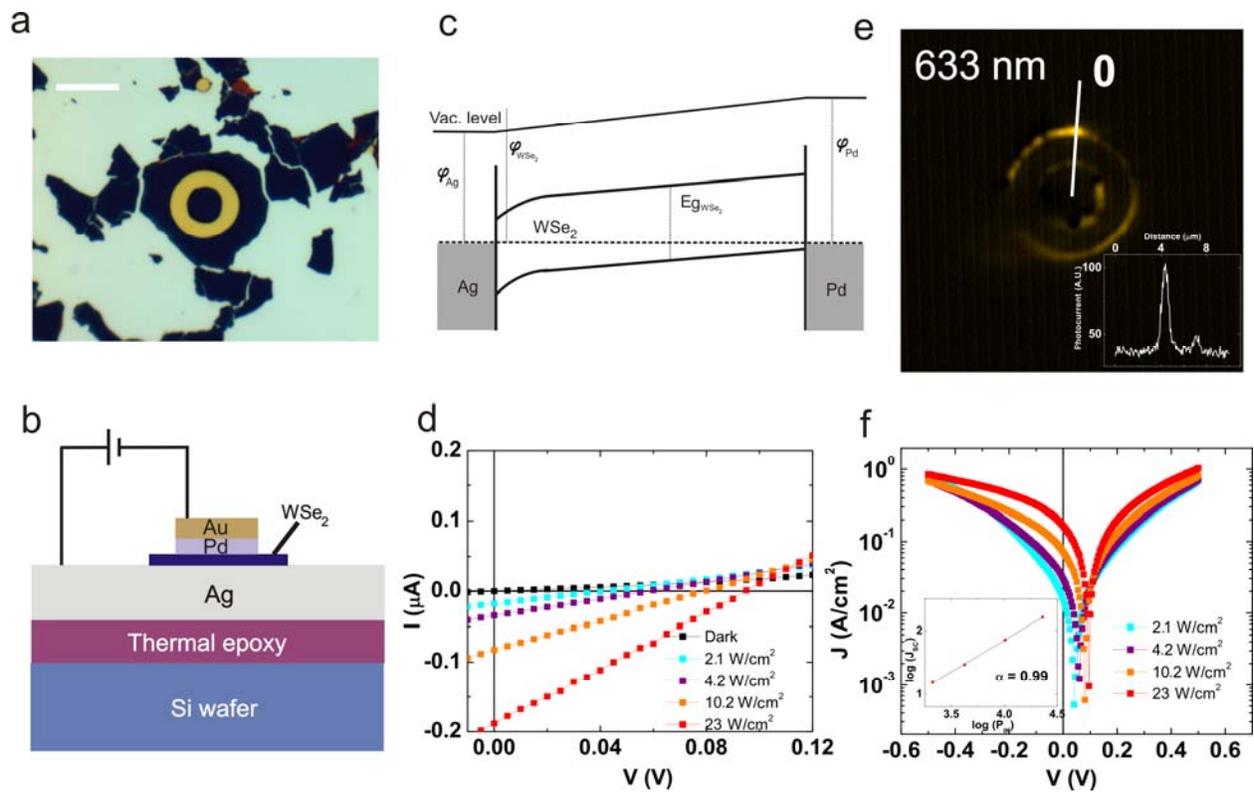

**Figure 4: Device structure and characteristics a.** Optical micrograph of a representative device comprising of 13 nm WSe$_2$ on Ag with a Pd/Au ring electrode on the top (Scale bar = 10 μm) **b.** Schematic representation of side view of the device in **(a). c.** Schematic band diagram showing Schottky contact on the Ag side and ohmic contact on the Pd side with a depleted WSe$_2$ in between. **d.** Current-voltage characteristics of a representative device (13 nm WSe$_2$/Ag) under dark and broadband white light illumination from a Hg-vapor lamp source. **e.** Spatially varying photocurrent map of the device acquired at 16 μW incident power. Inset shows the line profile of photocurrent magnitude along the white line in the map. The photocurrent profile suggests carrier diffusion length of ~ 1.5 μm. **f.** Current density vs voltage (J-V) curves estimated based on the active area determine from (**e**) and I-V plots from (**d**). Inset shows circuit current density proportional to input power with an exponent α = 0.99.

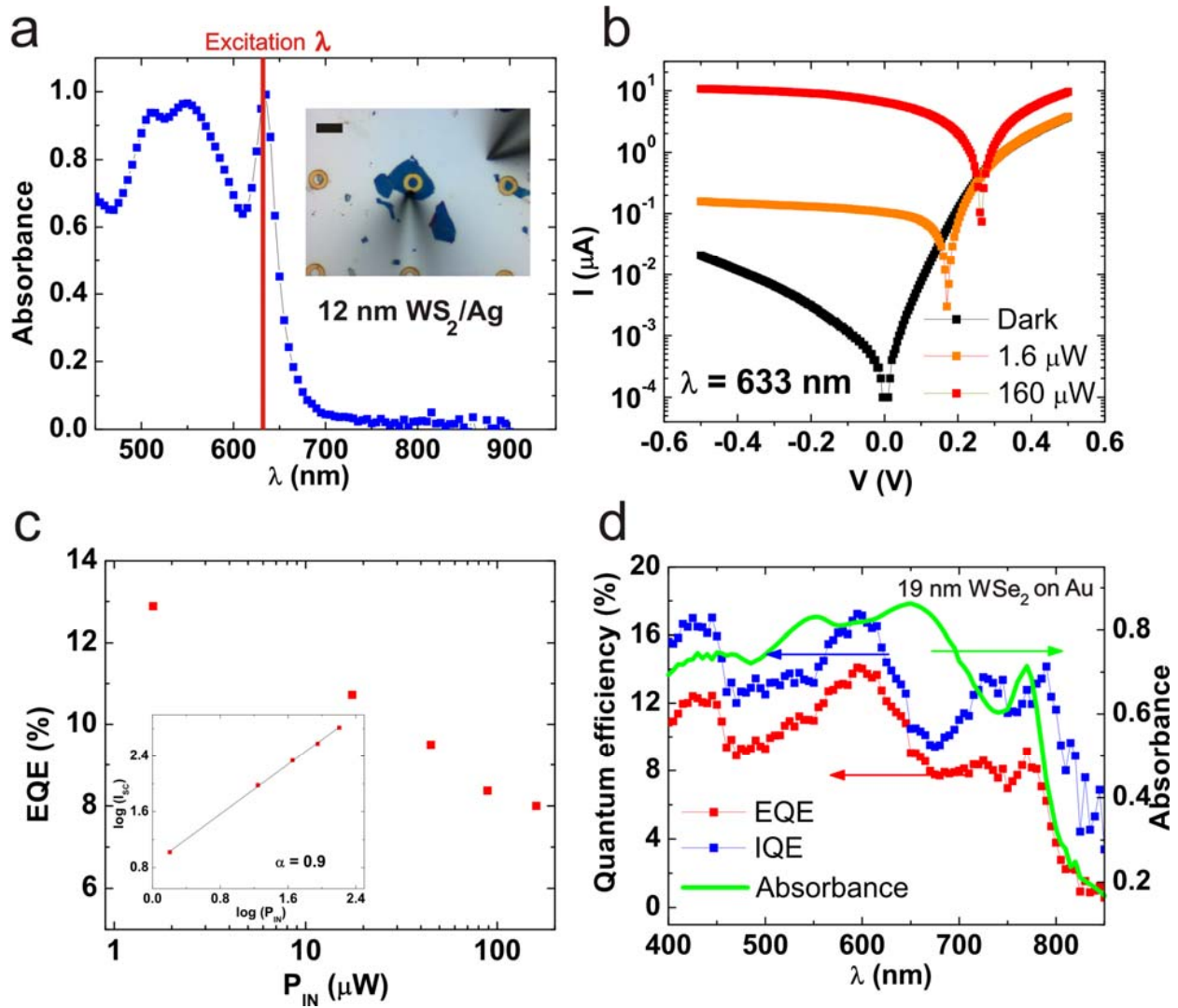

**Figure 5. Monochromatic illumination and external quantum efficiency: a.** Absorbance spectrum of 12nm $WS_2$/Ag stack. A near-unity absorbance is observed at the primary exciton peak. The red line denotes the 633 nm excitation wavelength. Inset shows the optical micrograph of the device along with electrical probes (Scale bar = 10 μm). **b.** I-V characteristics of the device in **(a)** with the 633 nm laser focused on a photocurrent producing spot. **c.** Power dependence of EQE for 633 nm incident laser. Inset shows an exponent of ~0.9 for power dependence of photocurrent for this device. **d.** EQE (red) and IQE (blue) spectra for a 19 nm

WSe$_2$ on Au device showing power generation across the absorption spectrum (green). The laser power is ~ 1 µW for 650 nm with about 10% variation across the spectrum.

**References:**

1.      Sukhovatkin, V.; Hinds, S.; Brzozowski, L.; Sargent, E. H. *Science* **2009,** 324, 1542-1544.
2.      Konstantatos, G.; Howard, I.; Fischer, A.; Hoogland, S.; Clifford, J.; Klem, E.; Levina, L.; Sargent, E. H. *Nature* **2006,** 442, 180-183.
3.      Peumans, P.; Bulović, V.; Forrest, S. *Appl. Phys. Lett.* **2000,** 76, 2650-2652.
4.      Yang, Y.; Zheng, Y.; Cao, W.; Titov, A.; Hyvonen, J.; Manders, J. R.; Xue, J.; Holloway, P. H.; Qian, L. *Nat. Photon.* **2015**, 10.1038/nphoton.2015.36.
5.      Polman, A.; Knight, M.; Garnett, E. C.; Ehrler, B.; Sinke, W. C. *Science* **2016,** 352.
6.      Atwater, H. A.; Polman, A. *Nat. Mater.* **2010,** 9, 205-213.
7.      Polman, A.; Atwater, H. A. *Nat. Mater.* **2012,** 11, 174-177.
8.      Vandamme, N.; Chen, H. L.; Gaucher, A.; Behaghel, B.; Lemaitre, A.; Cattoni, A.; Dupuis, C.; Bardou, N.; Guillemoles, J. F.; Collin, S. *IEEE Journal of Photovoltaics* **2015,** 5, 565-570.
9.      Kelzenberg, M. D.; Boettcher, S. W.; Petykiewicz, J. A.; Turner-Evans, D. B.; Putnam, M. C.; Warren, E. L.; Spurgeon, J. M.; Briggs, R. M.; Lewis, N. S.; Atwater, H. A. *Nat. Mater.* **2010,** 9, 239-244.
10.     Garnett, E.; Yang, P. *Nano Lett.* **2010,** 10, 1082-1087.
11.     Spinelli, P.; Verschuuren, M. A.; Polman, A. *Nat. Commun.* **2012,** 3, 692.
12.     Bermel, P.; Luo, C.; Zeng, L.; Kimerling, L. C.; Joannopoulos, J. D. *Optics Express* **2007,** 15, 16986-17000.
13.     Zeng, L.; Yi, Y.; Hong, C.; Liu, J.; Feng, N.; Duan, X.; Kimerling, L. C.; Alamariu, B. A. *Appl. Phys. Lett.* **2006,** 89, 111111.
14.     Ferry, V. E.; Sweatlock, L. A.; Pacifici, D.; Atwater, H. A. *Nano Lett.* **2008,** 8, 4391-4397.
15.     Ferry, V. E.; Munday, J. N.; Atwater, H. A. *Adv. Mater.* **2010,** 22, 4794-4808.
16.     Jariwala, D.; Sangwan, V. K.; Lauhon, L. J.; Marks, T. J.; Hersam, M. C. *ACS Nano* **2014,** 8, 1102–1120.
17.     Britnell, L.; Ribeiro, R. M.; Eckmann, A.; Jalil, R.; Belle, B. D.; Mishchenko, A.; Kim, Y.-J.; Gorbachev, R. V.; Georgiou, T.; Morozov, S. V.; Grigorenko, A. N.; Geim, A. K.; Casiraghi, C.; Neto, A. H. C.; Novoselov, K. S. *Science* **2013,** 340, 1311-1314
18.     Yu, W. J.; Liu, Y.; Zhou, H.; Yin, A.; Li, Z.; Huang, Y.; Duan, X. *Nat. Nanotechnol.* **2013,** 8, 952–958.
19.     Pospischil, A.; Furchi, M. M.; Mueller, T. *Nat. Nanotechnol.* **2014,** 9, 257-261.
20.     Jariwala, D.; Howell, S. L.; Chen, K.-S.; Kang, J.; Sangwan, V. K.; Filippone, S. A.; Turrisi, R.; Marks, T. J.; Lauhon, L. J.; Hersam, M. C. *Nano Lett.* **2016,** 16, 497–503.
21.     Baugher, B. W.; Churchill, H. O.; Yafang, Y.; Jarillo-Herrero, P. *Nat. Nanotechnol.* **2014,** 9, 262-267.
22.     Liu, X.; Galfsky, T.; Sun, Z.; Xia, F.; Lin, E.-c.; Lee, Y.-H.; Kéna-Cohen, S.; Menon, V. M. *Nat. Photon.* **2015,** 9, 30-34.
23.     Wang, W.; Klots, A.; Yang, Y.; Li, W.; Kravchenko, I. I.; Briggs, D. P.; Bolotin, K. I.; Valentine, J. *Appl. Phys. Lett.* **2015,** 106, 181104.

24. Lien, D.-H.; Kang, J. S.; Amani, M.; Chen, K.; Tosun, M.; Wang, H.-P.; Roy, T.; Eggleston, M. S.; Wu, M. C.; Dubey, M.; Lee, S.-C.; He, J.-H.; Javey, A. *Nano Lett.* **2015,** 15, 1356-1361.
25. Piper, J. R.; Fan, S. *ACS Photonics* **2016,** 3, 571-577.
26. Bahauddin, S. M.; Robatjazi, H.; Thomann, I. *ACS Photonics* **2016,** 3, 853-862.
27. Callahan, D. M.; Munday, J. N.; Atwater, H. A. *Nano Lett.* **2012,** 12, 214-218.
28. Yu, Z.; Raman, A.; Fan, S. *Proc. Nat. Acad. Sci. USA* **2010,** 107, 17491-17496.
29. Kats, M. A.; Blanchard, R.; Genevet, P.; Capasso, F. *Nat. Mater.* **2013,** 12, 20-24.
30. Grigorieva, I. V.; Geim, A. K. *Nature* **2013,** 499, 419-425.
31. McMorrow, J. J.; Walker, A. R.; Sangwan, V. K.; Jariwala, D.; Hoffman, E.; Everaerts, K.; Facchetti, A.; Hersam, M. C.; Marks, T. J. *ACS Appl. Mater. Interfaces* **2015,** 7, 26360-26366.
32. Novoselov, K. S.; Jiang, D.; Schedin, F.; Booth, T. J.; Khotkevich, V. V.; Morozov, S. V.; Geim, A. K. *Proc. Nat. Acad. Sci. USA* **2005,** 102, 10451-10453.
33. Wang, Q. H.; Kalantar-Zadeh, K.; Kis, A.; Coleman, J. N.; Strano, M. S. *Nat. Nanotechnol.* **2012,** 7, 699-712.
34. Li, Y.; Chernikov, A.; Zhang, X.; Rigosi, A.; Hill, H. M.; van der Zande, A. M.; Chenet, D. A.; Shih, E.-M.; Hone, J.; Heinz, T. F. *Phys. Rev. B* **2014,** 90, 205422.
35. Beal, A. R.; Liang, W. Y.; Hughes, H. P. *Journal of Physics C: Solid State Physics* **1976,** 9, 2449.
36. Beal, A. R.; Hughes, H. P. *Journal of Physics C: Solid State Physics* **1979,** 12, 881.
37. Nayak, P. K.; Yeh, C.-H.; Chen, Y.-C.; Chiu, P.-W. *ACS Appl. Mater. Interfaces* **2014,** 6, 16020-16026.
38. Zhao, W.; Ghorannevis, Z.; Chu, L.; Toh, M.; Kloc, C.; Tan, P.-H.; Eda, G. *ACS Nano* **2013,** 7, 791-797.
39. Grandidier, J.; Weitekamp, R. A.; Deceglie, M. G.; Callahan, D. M.; Battaglia, C.; Bukowsky, C. R.; Ballif, C.; Grubbs, R. H.; Atwater, H. A. *Phys. Status Solidi A* **2013,** 210, 255-260.
40. Komsa, H.-P.; Krasheninnikov, A. V. *Phys. Rev. B* **2012,** 86, 241201.
41. Lakhwani, G.; Rao, A.; Friend, R. H. *Annu. Rev. Phys. Chem.* **2014,** 65, 557-581.
42. Lee, C.-H.; Lee, G.-H.; van der Zande, A. M.; Chen, W.; Li, Y.; Han, M.; Cui, X.; Arefe, G.; Nuckolls, C.; Heinz, T. F.; Guo, J.; Hone, J.; Kim, P. *Nat. Nanotechnol.* **2014,** 9, 676-681
43. Amani, M.; Lien, D.-H.; Kiriya, D.; Xiao, J.; Azcatl, A.; Noh, J.; Madhvapathy, S. R.; Addou, R.; KC, S.; Dubey, M.; Cho, K.; Wallace, R. M.; Lee, S.-C.; He, J.-H.; Ager, J. W.; Zhang, X.; Yablonovitch, E.; Javey, A. *Science* **2015,** 350, 1065-1068.
44. Kang, K.; Xie, S.; Huang, L.; Han, Y.; Huang, P. Y.; Mak, K. F.; Kim, C.-J.; Muller, D.; Park, J. *Nature* **2015,** 520, 656-660.

Supporting Information

# Near-Unity Absorption in Van der Waals Semiconductors for Ultrathin Optoelectronics

*Deep Jariwala,[1,2] Artur R. Davoyan,[1,2,3] Giulia Tagliabue,[1,4] Michelle C. Sherrott,[1,2] Joeson Wong[1] and Harry A. Atwater[1,2,3,4]\**

[1]Department of Applied Physics and Materials Science, California Institute of Technology, Pasadena, CA-91125, USA

[2]Resnick Sustainability Institute, California Institute of Technology, Pasadena, CA-91125, USA

[3]Kavli Nanoscience Institute, California Institute of Technology, Pasadena, CA-91125, USA

[4]Joint Center for Artificial Photosynthesis, California Institute of Technology, Pasadena, CA-91125, USA

*Corresponding author: haa@caltech.edu

**S1. Absorbance calculations:**

Absorption in ultrathin TMDCs on metals was calculated using the transfer matrix method.[1] The optical constants used for the calculation are the bulk crystal values for each TMDCs from previously published reports.[2,3] Permittivities of Ag and Au were taken from Jhonson and Christie[4].

The maximum integrated absorbance in ultrathin TMDCs on a reflective metallic substrate is only achieved for a critical TMDC thickness which lies somewhere between 12-15 nm. Below this critical thickness, the absorption in the red part of the spectrum is reduced due to high reflection from the underlying metal. Above this thickness, the absorption in the blue part of the spectrum is reduced due to increased reflection from the TMDC owing to the mismatch in refractive index between air and the TMDC. Figure 2 c-d in the manuscript can be seen for experimental verification. The coupling of TMDCs with reflective metals is crucial for this resonantly enhanced absorption to occur. In case of a free standing $WSe_2$ of similar thickness, the total absorption is far lesser approaching ~40% between 6-12 nm thickness before falling down again due to increased reflection owing to index mismatch at the air /TMDC interface (Figure S1).

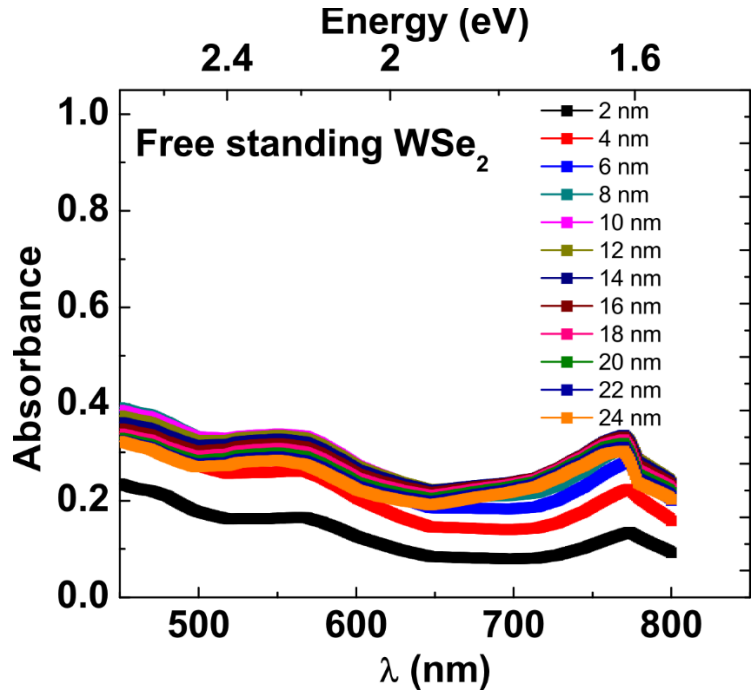

Figure S1: Calculated absorption in free standing $WSe_2$ with varying thickness. The total absorption increases with increasing thickness upto 12-14 nm and approaches ~40% following which it steadily drops down with further increase in the thickness.

## S2. Absorption with Au back reflectors:

Figure S3 below shows calculated and measured absorption spectra for varying thicknesses of $WSe_2$ and $WS_2$ on template stripped Au back reflectors. A good qualitative and quantitative agreement between the measured and calculated spectra is apparent once again. However in Au, interband absorption starts dominating below 550 nm in wavelength. Therefore, the useful absorption in the TMDC layer (dashed lines) drastically reduces at $\lambda < 500$ nm. Ag back reflector is thus more suitable from an optical standpoint of maximizing useful absorption.

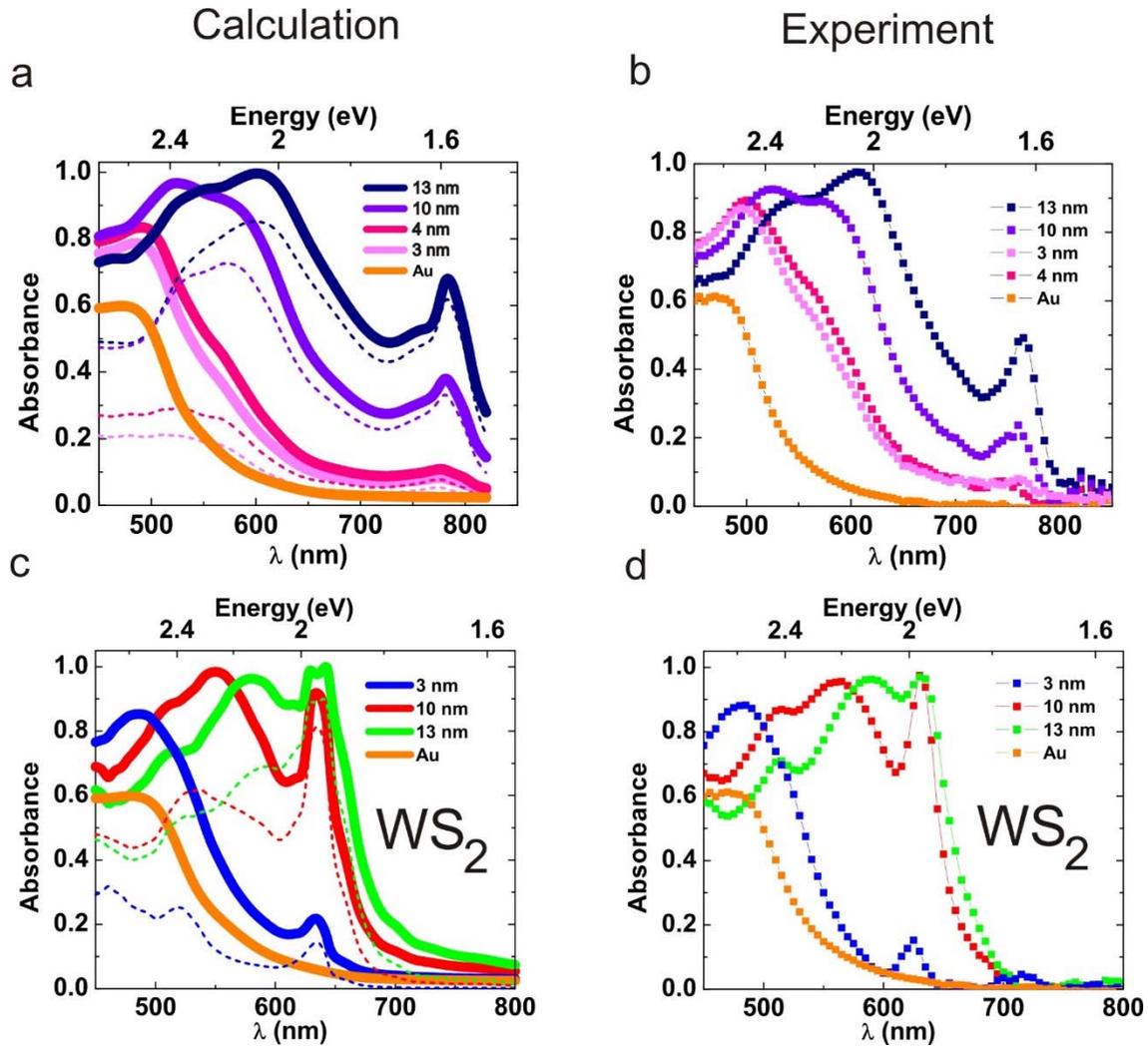

Figure S3: Calculated (left) and measured (right) absorption spectra of WSe$_2$ on Au (a-b) and WS$_2$ on Au(c-d).

**S3. Estimation of minority carrier diffusion length.**

Minority carrier diffusion length can be estimated from the spatial photocurrent profile. In cases where the diffusion length is larger than the spot size, a single exponential model $I = I_0 \, exp(-x/L_D)$ where I is the photocurrent, $I_0$ is the peak photocurrent, $x$ is the distance and $L_D$ is the diffusion length can explain the photocurrent profile and provide an estimate of diffusion length. Considering that we are using a 633 nm laser for spatially resolving the photocurrent, the diffraction limited resolution (given by 0.61λ/N.A.), where N.A. is the numerical aperture of the objective, = 0.8 in our case) is ~500 nm in our measurement. Based on that, we can fit our photocurrent data to the above exponential decay equation and extract minority carrier diffusion lengths. We have estimated diffusion length varying from about 1.35 μm in Figure S4 a to about 3 μm in Figure S4 b.

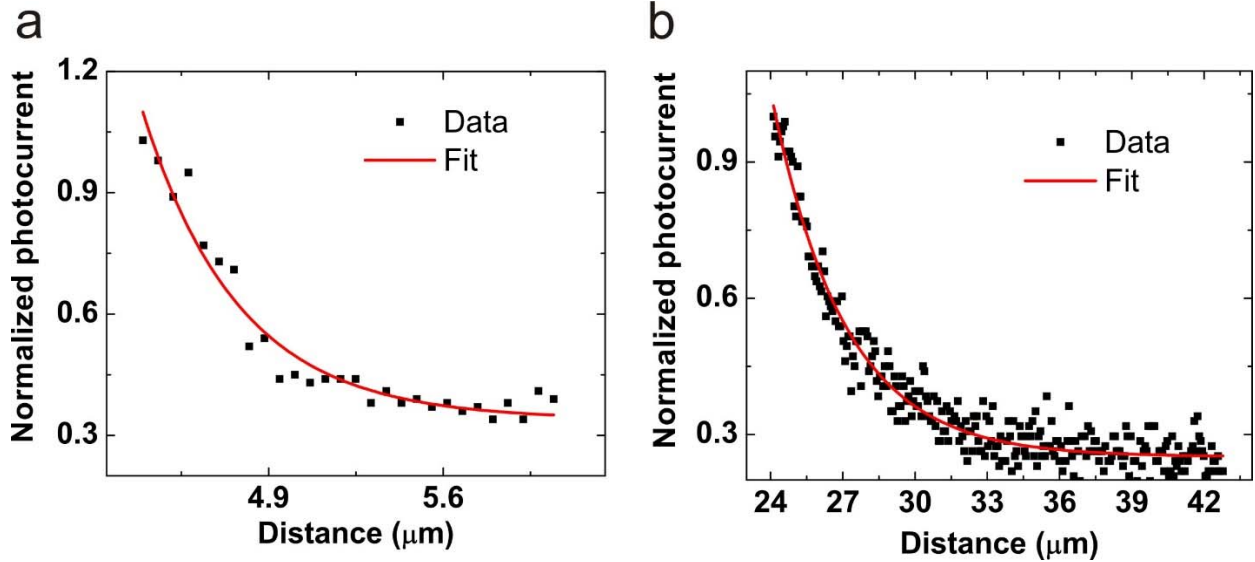

Figure S4. a. Photocurrent profile of device shown in Figure 3 e of the manuscript with a minority carrier diffusion length of ∼1.35 μm b. Photocurrent profile of another representative device (19 nm $WSe_2$ on Au) shown in Figure 4 e of the manuscript with a minority carrier diffusion length of ∼3 μm.

## S4. Absorbance and EQE measurements:

A home-built optical set-up was used for both the absorption and EQE measurements. A supercontinuum laser (Fianium) coupled to a monochromator was used to provide the monochromated incident light. The collimated beam was focused onto the sample with a long working distance (NA = 0.55) 50x objective in order to achieve nearly normal illumination. The reflection spectrum was measured with a Si photodetector. Low noise signals were obtained by using a chopper and a lock-in amplifier. The measured reflection signal was then normalized to the reflection from a silver mirror (Thorlabs) in order to obtain the absolute reflection spectrum, $R(\lambda)$. In the absence of any transmission, the absorption spectrum can be obtained as $A(\lambda) = 1-R(\lambda)$.

The same illumination configuration was used for the EQE measurements. The photocurrent signal produced by the TMDC device was measured at each wavelength by mean of the chopper and lock-in amplifier. In addition, the power spectrum incident on the sample was later measured by placing the Si photodetector in the same position as the sample.

During all measurements, a small fraction of the illumination beam is deviated onto an optical fiber and sent to a second lock-in amplifier, also driven at the same frequency of the chopper. This reference signal is used to account for fluctuations of the illuminating beam over time enabling accurate normalization of the reflection and photocurrent signals.

## S5. WS$_2$/WSe$_2$ heterojunctions:

Heterojunctions of few-layer WS$_2$/WSe$_2$ on Au substrates can also be fabricated by exfoliation and layer stacking using the dry transfer technique.[5] As compared to individual TMDC layers, heterojunctions show more enhanced, broadband absorption as shown below in Figure S5 below. Further since WS$_2$ and WSe$_2$ are known to form a type-II junction,[6] it is expected that the photocurrent collection efficiencies will also be enhanced in a optoelectronic device fabricated out of such heterojunctions with optimized layer thicknesses. Future work will involve fabricating metal ring contacts on such heterostructures and also fabricating graphene contacts to understand and enhance the carrier collection and open-circuit voltage in the resulting devices.

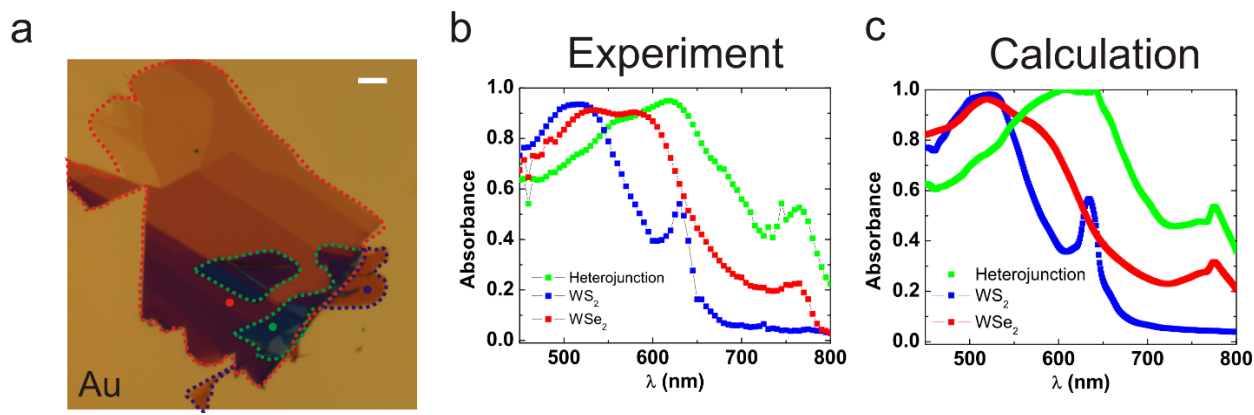

Figure S5: a. Optical micrograph of a WS$_2$/WSe$_2$/Au heterostructure of varying thicknesses (scale bar = 5 μm). The blue, red and green boundaries indicate the WS$_2$/Au, WSe$_2$/Au and WS$_2$/WSe$_2$/Au heterostructures regions respectively. The blue, red and green circles denote the spots from where the absorption spectra were acquired in b. b. Absorption spectra from the correspondingly colored circles in a. The layer thicknesses were measured using AFM. A clear increase in integrated absorption (area under the curve) is observed the case of heterojunction (green) vs individual WS$_2$ (blue) and WSe$_2$ (red) layers. c. Corresponding calculated spectra using the transfer matrix method in good agreement with the measurements in b.

**References:**

1. Born, M.; Wolf, E., *Principles of optics: electromagnetic theory of propagation, interference and diffraction of light*. Cambridge University Press: Cambridge, Unied Kingdom, 2000.
2. Li, Y.; Chernikov, A.; Zhang, X.; Rigosi, A.; Hill, H. M.; van der Zande, A. M.; Chenet, D. A.; Shih, E.-M.; Hone, J.; Heinz, T. F. *Phys. Rev. B* **2014,** 90, 205422.
3. Wilson, J. A.; Yoffe, A. D. *Advances in Physics* **1969,** 18, 193-335.
4. Johnson, P. B.; Christy, R. W. *Phys. Rev. B* **1972,** 6, 4370-4379.
5. Andres, C.-G.; Michele, B.; Rianda, M.; Vibhor, S.; Laurens, J.; Herre, S. J. v. d. Z.; Gary, A. S. *2D Mater.* **2014,** 1, 011002.
6. Wang, K.; Huang, B.; Tian, M.; Ceballos, F.; Lin, M.-W.; Mahjouri-Samani, M.; Boulesbaa, A.; Puretzky, A. A.; Rouleau, C. M.; Yoon, M.; Zhao, H.; Xiao, K.; Duscher, G.; Geohegan, D. B. *ACS Nano* **2016**, 10.1021/acsnano.6b01486.